\documentclass[vecphys]{svmult}

\usepackage{makeidx}         
\usepackage{graphicx}        
\usepackage{multicol}        
\usepackage[bottom]{footmisc}

\usepackage{mathrsfs,graphicx,bm,amsmath}
\usepackage{epsfig,color}
\usepackage{latexsym,amssymb,amsmath,bbm,graphicx,graphics,here,rotating}

\newcommand{\Tr}{\mathrm{Tr}}

\newcommand{\sigex}{\tilde{\sigma}}

\newcommand{\rex}{\tilde{r}}
\newcommand{\hpn}{\tilde{h}_\phi}
\newcommand{\Tn}{\tilde{T}}

\newcommand{\hF}{^{(F)}}

\newcommand{\lit}{^6\mathrm{Li}}
\newcommand{\kal}{^{40}\mathrm{K}}
\newcommand{\epm}{\epsilon_M}

\newcommand{\hpb}{\bar{h}_\phi}
\newcommand{\lpb}{\bar{\lambda}_\phi}

\newcommand{\OmB}{\bar{\Omega}_B}

\newcommand{\OmC}{\bar{\Omega}_C}
\newcommand{\OmM}{\bar{\Omega}_M}

\makeindex             

\begin{document}

\title*{Universality in the BCS - BEC Crossover in Cold Fermion Gases}
\author{S. Diehl }
\institute{Institut f{\"u}r Theoretische Physik,
Philosophenweg 16, 69120 Heidelberg, Germany
\texttt{S.Diehl@thphys.uni-heidelberg.de}}
%
%
\maketitle

\section{Introduction}

There are two cornerstones for the description of quantum condensation phenomena: Bose-Einstein condensation (BEC) in
bosonic systems and BCS-pairing for fermions. The underlying pictures for these phenomena are quite different -- BEC
is the macroscopic population of a single quantum state, while the BCS mechanism relies on the formation of
Cooper pairs. However, both phenomena share a decisive feature in common: they can be described as the spontaneous
breaking of the global symmetry of phase rotations, $U(1)$. Due to this similarity, it is plausible that the
two different scenarios sketched above are indeed connected by some smooth transition or crossover
\cite{Eagles69,ALeggett80,BNozieres85}.
In a simple physical picture, the position-space delocalized Cooper pairs characteristic for the BCS regime
undergo a localization process throughout the crossover, ending up as effectively pointlike bosonic particles or
strongly bound molecules in the BEC regime.

At this point ultracold atoms come into play. The presence of Feshbach resonances in fermionic gases such as
$\lit$ or $\kal$ offers the unique possibility to tune the
interaction strength between the atoms to arbitrary positive and negative values, thereby allowing for an experimental
implementation of the crossover \cite{Jin04,Ketterle04,ZGrimm04,Thomas04,Bourdel04,Strecker04}.

In this contribution \footnote{Talk given at the 2006 ECT* School ``Renormalization Group and Effective Field Theory
Approaches to Many-Body Systems'', Trento, Italy.}, we address the crossover problem based on an atom-molecule model which
is appropriate for a realistic description of the crossover in cold fermion gases. In sect. \ref{MMandFI} we summarize
the ingredients of the functional integral formalism developed in refs. \cite{Diehl:2005ae,Diehl:2005an}. In sect.
\ref{sec:Univ} we discuss universal aspects of the phase diagram encoded in the atom-molecule model, with special
emphasis on an additional form of crossover interpolating between universal narrow to broad resonance limits. While
the narrow resonance limit describing a situation with large
effective range can be solved exactly, broad resonances corresponding to pointlike interactions pose the challenge of
a strongly interacting field theory. We further investigate deviations from the broad resonance universality,
making contact to a recent experiment which focuses on the fraction of closed channel molecules throughout the
crossover. Our findings excellently agree with measurements. In sect. \ref{sec:Conclusion} we draw our conclusions.

\section{Functional Integral for the Crossover Problem}
\label{MMandFI}

\subsection{Microscopic Model}
\begin{figure}[t]
\begin{minipage}{\linewidth}
\begin{center}
\setlength{\unitlength}{1mm}
\begin{picture}(44,27)
   \put (0,0){
    \makebox(43,26){
\begin{picture}(43,26)
      \put(0,0){\epsfxsize40mm \epsffile{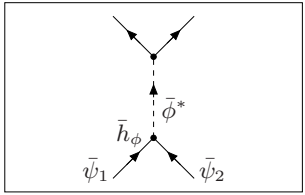}}
      \put(15,7){$\hpb$}
      \put(21,10){$\bar{\phi}^*$}
      \put(10.5,2){$\bar\psi_1$}
      \put(26,2){$\bar\psi_2$}
\end{picture}  }}
 \end{picture}
\end{center}
\vspace*{-1ex} \caption{Tree exchange of a molecule.}
\label{TreeEx}
\end{minipage}
\end{figure}

Our euclidean Yukawa-type bare microscopic action in position space is given by
\cite{BBKokkelmans,EEGriffin,AATimmermans,GGChen,WWStoofBos}:
\begin{eqnarray}\label{YukawaAction}
S  [\bar\psi,\bar\phi] &=& \int d\tau \int d^3x
\Big[\bar\psi^\dagger\big(\partial_\tau - \frac{\triangle}{2M} -\sigma\big)\bar\psi + \frac{\bar{\lambda}_{\psi,\Lambda}}{2}
(\bar\psi^\dagger\bar\psi)^2\\
&&\hspace{-0.12cm}+\bar{\phi}^*\big(\partial_\tau - \bar A_{\phi,0} \triangle + \bar{\nu}_\Lambda(B) - 2 \sigma\big)\bar{\phi}
-\bar{h}_{\phi,\Lambda}\Big(\bar{\phi}^*\bar\psi_1\bar\psi_2 - \bar{\phi}\bar\psi^*_1\bar\psi^*_2\Big)\Big].\nonumber
\end{eqnarray}
Here, $\bar\psi = (\bar\psi_1,\bar\psi_2)$ is the stable fermion field, i.e. a nonrelativistic 2-spinor whose
components represent the lowest hyperfine states $|1\rangle$ and $|2\rangle$ of a given atom species with mass $M$,
typically $\lit$ or $\kal$. The complex scalar $\bar \phi$ represents a bosonic
``closed channel'' or ``bare'' molecule. Both fields propagate non-relativistically, with their time evolution coefficients
normalized to one. They couple to a chemical potential $\sigma$, where the factor 2 for the molecules accounts for their
double atom number. The Feshbach resonant interaction is modelled by the tree exchange of a molecule
(cf. fig. \ref{TreeEx}). It is parameterized by the Feshbach coupling $\bar{h}_{\phi,\Lambda}$ and the
inverse classical molecule propagator. This allows for the
description of nonlocal interactions as detailed in sect. \ref{sec:decoupling}. The crossover from BCS to BEC is
described by the mass-like (bare) detuning parameter $\bar\nu_\Lambda(B)$ which depends on the magnetic field.
The classical gradient coefficient $\bar A_{\phi,0}$ can be related to an effective range by
$r_s = 2 \bar A_{\phi,0}/ \bar h_\phi^2$ \cite{Diehl:2006}. Further, a pointlike non-resonant
four-fermion interaction in the open or background channel is implemented by the coupling $\bar{\lambda}_{\psi,\Lambda}$.

\begin{figure}[t]
\begin{minipage}{\linewidth}
\begin{center}
\setlength{\unitlength}{1mm}
\begin{picture}(72,47)
      \put (0,0){
     \makebox(72,45){
 \begin{picture}(72,45)
      \put(0,0){\epsfxsize70mm \epsffile{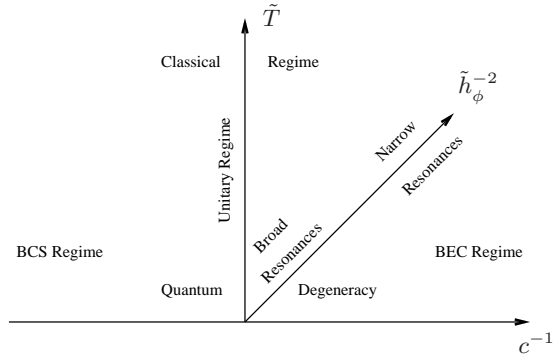}}
      \put(68,-3.5){$c^{-1}$}
      \put(60,31){$\hpn^{-2}$}
      \put(34,40){$\Tn$}
\end{picture}  }}
   \end{picture}
\end{center}
\vspace*{-1ex} \caption{Cube of scales for the crossover problem. }
\label{CubeOfScales}
\end{minipage}
\end{figure}

The classical action features bare parameters which need to be related to actual microphysical observables by an
ultraviolet renormalization procedure. This maps $\{\bar\nu_\Lambda(B), \bar{h}_{\phi,\Lambda}, \bar{\lambda}_{\psi,\Lambda}\}
\to \{\bar\nu(B), \hpb, \bar{\lambda}_{\psi}\}$. Furthermore, the system is more
efficiently parameterized by trading the physical detuning $\bar\nu(B)$ for an in-medium scattering length \footnote{The
physical scattering length $a$ obtains as $a= a(\sigma =0)$. The in-medium scattering length is, however, the
appropriate quantity to characterize the universal ground state properties of the system, \cite{Diehl:2006}. For broad
resonances, $a(\sigma) \approx a$.}
\begin{eqnarray}\label{InMediumScattLength}
a (\sigma)  =- \frac{M \hpb^2}{4\pi(\bar{\nu}(B) - 2\sigma)} + a_{bg}, \quad a_{bg} =\frac{M \bar{\lambda}_{\psi}}{4\pi}.
\end{eqnarray}
The action can be brought into a dimensionless scaling form where all fields, couplings and coordinates are measured in
units of the Fermi momentum $k_F := (3\pi^2 n)$ ($n$ the particle density which sets a natural scale for
thermodynamics) or the Fermi energy $\epsilon_F = k_F^2/(2M)$. This defines dimensionless parameters
\begin{eqnarray}
c^{-1} = (a(\sigma) k_F)^{-1}, \, \hpn= 2M k_F^{-1/2} \hpb, \, \tilde{\lambda}_{\psi} =
2Mk_F\bar{\lambda}_{\psi}, \, \tilde A_{\phi,0} = 2M A_{\phi,0}.
\end{eqnarray}
The part of the parameter space spanned by the variables which we focus on here is shown in a ``cube of scales'',
fig. \ref{CubeOfScales} \footnote{We omit a fourth axis for $a_{bg} k_F$. This coupling will, however, play a role
in sect. \ref{sec:DeviatUniv}.}. The crossover from BCS- to BEC-type physics takes place on the $c^{-1}$ - axis, the
coupling $c$ parameterizing the strength of the fermionic interaction. The values
of the Feshbach coupling $\hpn$ define an additional axis which gives rise to a further crossover from narrow (small
$\hpn$) to broad (large $\hpn$) resonances. The most challenging region is centered around the origin of the cube
of scales, where we deal with a strongly interacting non-relativistic quantum field theory.

\subsection{Functional Integral}

We quantize the microscopic action by means of a functional integral for both fermionic and bosonic fields. The effective
action can be represented as
\begin{eqnarray}\label{EffActStart}
\Gamma[\bar\Psi ,\bar\Phi] =-\log \int \mathcal{D}\delta\hat{\bar\Psi}\mathcal D \delta\hat{\bar\Phi} \exp\big( -
S  [\hat{\bar\Psi}, \hat{\bar\Phi} ]
+  J_\phi^T\delta\hat{\bar\Phi}+  J_\psi^T\delta\hat{\bar\Psi}\big)
\end{eqnarray}
with fermionic and bosonic Nambu fields $\hat{\bar\Psi} = (\hat{\bar\psi},\hat{\bar\psi}^*)^T, \hat{\bar\Phi}
 = (\hat{\bar\phi},\hat{\bar\phi}^*)^T$,
and corresponding sources $J_\psi, J_\phi$ which are Grassmann (complex) valued. The effective action is a functional
of the ``classical fields'' (field expectation values), such that it is favorable to decompose the fields
into a classical part $\bar\Psi = \langle \hat{\bar\Psi}\rangle, \bar\Phi = \langle \hat{\bar\Phi}\rangle$ and a
fluctuation $\delta \hat{\bar\Psi}, \delta \hat{\bar \Phi}$, i.e. $\hat{\bar\Psi} = \bar\Psi + \delta \hat{\bar\Psi},
\hat{\bar\Phi} = \bar\Phi + \delta \hat{\bar\Phi}$.

\begin{figure}
\begin{center}
\setlength{\unitlength}{1mm}
\begin{picture}(95,60)
      \put (0,0){
     \makebox(90,57){
     \begin{picture}(90,57)
      \put(0,0){\epsfxsize90mm \epsffile{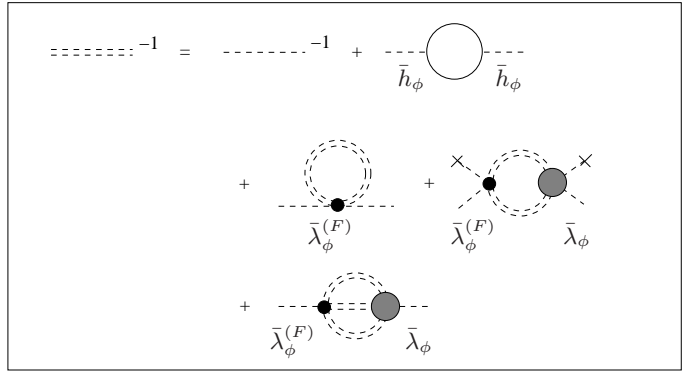}}
      \put(35,3){$\lpb\hF$}
      \put(53,3){$\lpb$}
      \put(40,17){$\lpb\hF$}
      \put(59,17){$\lpb\hF$}
      \put(74,17){$\lpb$}
      \put(52,38){$\hpb$}
      \put(65,38){$\hpb$}
     \end{picture}
      }}
   \end{picture}
\end{center}
\vspace*{-1ex}
\caption{Schwinger-Dyson equation for the inverse molecule propagator (double dashed line).
The first two terms on the rhs denote the ``mean field inverse propagator'' after integrating out
the fermionic fluctuations with a dashed line for the classical inverse molecule propagator and a solid line for the fermion
propagator. The terms in the second line account for the one-loop molecule fluctuations. Here the small full circle
$\lpb\hF$ is the molecule self-interaction induced by the fermion fluctuations. The large shaded circles
represent the full four-boson coupling $\lambda_\phi$. The two-loop term is neglected in our calculations.}
\label{SDEGraphic}
\end{figure}

To illustrate our evaluation strategy, we concentrate on $\bar \lambda_\psi = 0$ here. Then the action (\ref{YukawaAction}) is quadratic in the fermions, such
that we can integrate them out in one step \footnote{Additionally, the equation of motion for the fermions be satisfied,
i.e. $J_\psi=0$.}. Inserting the physical fermion field expectation value $\bar\Psi=0$ (Pauli's principle), this yields
a purely bosonic theory,
\begin{eqnarray}\label{PurelyBosonic}
\Gamma[\bar\Psi=0,\bar\Phi] =
 - \log \int \mathcal D \delta \hat{\bar\Phi} \exp\big( - \bar S[ \delta\hat{\bar\Phi} + \bar\Phi]
 +  J_\phi^T\delta\hat{\bar\Phi}\big)
\end{eqnarray}
with an intermediate action $\bar S$ depending on the fluctuating field $\hat{\bar\Phi} = \bar\Phi + \delta\hat{\bar\Phi}$,
and given by the exact expression
\begin{eqnarray}\label{IntermediateAct}
\bar S [\hat{\bar\Phi}] = S_\phi^{(cl)}[\hat{\bar\Phi}] - \frac{1}{2}\log\det S  ^{(\psi\psi)}[\hat{\bar\Phi}] =
S_\phi^{(cl)}[\hat{\bar\Phi}] - \frac{1}{2}\Tr \log S  ^{(\psi\psi)}[\hat{\bar\Phi}]
\end{eqnarray}
where $S  ^{(\psi\psi)}$ denotes the second variation w.r.t. the fermion fields.
The further evaluation has to deal with the remaining bosonic functional integral. For this purpose we
study a set of Schwinger-Dyson equations (SDE) for various bosonic couplings. The SDE for the full effective boson
propagator is displayed graphically in fig. \ref{SDEGraphic}. We truncate the SDE at the one-loop level. The loop
expansion should be reliable away from the critical line, but might become questionable close to the phase transition
where strong bosonic fluctuation effects occur. Further we use a combined vertex plus derivative expansion. We
work up to fourth order in the boson field for the vertex expansion. This corresponds to the analysis of an effective
$\phi^4$-theory. The derivative expansion keeps the leading order terms -- we are interested in the low energy
theory here. The diagrammatic structure then leads to the
familiar form of the frequency and momentum dependence of the effective inverse boson propagator for nonrelativistic
bosons, with coefficients determined from the solution of the Schwinger-Dyson equations. For a detailed
presentation of the technical issues, cf. \cite{Diehl:2006}. Importantly, our scheme allows for a self-consistent
determination of the bosonic couplings entering our truncation.
An alternative evaluation strategy which we currently investigate \cite{SDHGJPCW} uses the Functional
Renormalization Group for the effective action \cite{CWRG,Tetradis}. Here the mode elimination is performed
simultaneously for fermionic and bosonic degrees of freedom.

\subsection{Field Theoretical Construction}
\label{FtConstr}

The formulation of the crossover problem in terms of a functional integral allows to both systematically construct the
equation of state and to classify the thermodynamic phases from a symmetry consideration. Furthermore, an appropriately
performed vacuum limit of the effective action fixes the microphysical observables and provides for an ultraviolet
renormalization procedure.

\subsubsection{Equation of State}

The particle number can be obtained as the conserved charge of our nonrelativistic, $U(1)$ symmetric theory. Performing
the Noether construction for the effective action in the form (\ref{PurelyBosonic}), we are left with an equation of state
formulated in terms of couplings which can be readily extracted from our system of SDEs. Furthermore, this procedure
introduces the concept of dressed fields $\phi := Z_\phi \bar \phi$ in a natural way. Here $\bar \phi$ represents the
bare molecule field appearing in the microscopic theory and $Z_\phi$ the wave function renormalization as extracted from
the frequency dependence of the effective inverse boson propagator. Introducing a rescaling transformation which
normalizes the coefficient of the bosonic frequency term to one (standard time evolution), the particle density
manifestly reduces to an equation of state for effective bosons with atom number 2 in the BEC regime. This is further
explained in sect. \ref{sec:Univ}.

\subsubsection{Classification of the Thermodynamic Phases}

We can use the effective action formalism to classify the phases of the system according to the symmetries of the ground
state (thermodynamic equilibrium). In a homogeneous situation, we can consider the effective potential $\tilde u$
($\tilde u = \Gamma/V$ in the homogeneous limit) which can only depend on the invariant $\rho = \phi^*\phi$.
The field equation reads
\begin{eqnarray}\label{PotFieldEq}
\frac{\partial\tilde u}{\partial \phi^*}= \frac{\partial \tilde u}{\partial \rho}(\rho) \cdot \phi =
m_\phi^2(\rho) \cdot \phi = 0.
\end{eqnarray}
We have defined a bosonic mass term $m_\phi^2$ as the $\rho$-derivative of the effective potential. This equation can be
used to classify the phases of the system ($\rho_0$ denotes the solution of eq. (\ref{PotFieldEq})):
\begin{eqnarray}\label{CharPhases}
\mathrm{Symmetric\,\,phase:} && \rho_0 =0, \quad m_\phi^2 > 0,\\\nonumber
\mathrm{Symmetry\,\, broken\,\, phase:} &&\rho_0  > 0, \quad m_\phi^2 = 0,\\\nonumber
\mathrm{Phase\,\, transition:} && \rho_0 = 0,\quad m_\phi^2 = 0.
\end{eqnarray}
In the symmetric phase (SYM), we deal with a normal gas where there is no condensate, $\rho_0=0$. The symmetry broken
phase (SSB) is characterized by a nonvanishing field expectation value. Eq. (\ref{PotFieldEq}) implies the vanishing of
the mass term $m_\phi^2$. The massless mode reflects Goldstone's theorem and is responsible for superfluidity. In the
nonrelativistic bosonic theory, the Goldstone boson additionally manifests itself in a linear dispersion relation
$\omega = v_s |q|$. Our derivation recovers this relation with a speed of sound parameterized by $\tilde v_s =
\sqrt{2 A_\phi \lambda_\phi \rho_0}, \tilde v_s = 2M v_s/ k_F$. The phase transition is characterized by the
simultaneous vanishing of the mass term and the condensate. The additional constraint allows to solve eq.
(\ref{PotFieldEq}) for the critical temperature.

\subsubsection{Vacuum Limit}
\label{sec:VacLim}

In order to make contact to experiments the model parameters need to be related to observable quantities.
The effective action (\ref{EffActStart}) is formulated for arbitrary temperature $T$ and particle density $n$. However,
we can project on the physical vacuum in an appropriately performed low density limit,
\begin{eqnarray}\label{GammaVac}
\Gamma(vac) = \lim\limits_{k_F \to 0}\Gamma |_{\Tn >\Tn_c}.
\end{eqnarray}
This prescription dilutes the system by sending the mean interparticle spacing $k_F^{-1}\to \infty$. At the same time,
the dimensionless temperature $\Tn = 2M T/k_F^2$ is kept above criticality to ensure the absence of collective effects.
The absolute temperature obviously goes to zero $\sim k_F^2$. Eq. (\ref{GammaVac}) implies the following constraints
($\sigma_A$ is the ``chemical potential'' for the fermions in vacuum, $\bar m_\phi^2 = Z_\phi m_\phi^2$ the
unrenormalized bosonic mass term),
\begin{eqnarray}\label{2bdConstraint}
\mathrm{BCS}\,\, (a^{-1} <0)  &:&  \quad \sigma_A = 0 , \quad \bar m_\phi^2 > 0,\\\nonumber
\mathrm{BEC}\,\, (a^{-1} >0)  &:&  \quad \sigma_A < 0 , \quad \bar m_\phi^2 = 0,\\\nonumber
\mathrm{Resonance}\,\, (a^{-1}= 0)  &:&  \quad \sigma_A = 0 , \quad \bar m_\phi^2 = 0.
\end{eqnarray}
These constraints have a simple physical interpretation. The inverse fermion and effective boson (Minkowski-)
propagators in the absence of spontaneous symmetry breaking are given by \footnote{IR divergences in $Z_\phi,\bar A_\phi$
for $\sigma_A \to 0$ question the validity of the derivative expansion for the boson propagator on the BCS side. On the
other hand, physical combinations such as $A_\phi/Z_\phi$ or the scattering length remain finite for $\sigma_A \to 0$.
Omitting the derivative expansion, the effective boson propagator is free of IR divergences \cite{Diehl:2006},
such that the statements made here remain valid.}
\begin{eqnarray}
P_F = \mathrm i \omega + q^2/(2M) - \sigma , \qquad P_\phi = \mathrm i Z_\phi \omega + \bar A_\phi q^2 + \bar m_\phi^2.
\end{eqnarray}
Eq. (\ref{2bdConstraint}) implies massless fermions and massive bosons on the BCS side -- the propagation
of the ``molecules'' is hampered on the BCS side, and the fermionic atoms are the propagating degrees of freedom.
The situation is reversed on the BEC side. The ground state is a stable molecule, and the fermionic chemical potential
can be interpreted as half the binding energy of a molecule, $\epm = 2 \sigma_A$ \cite{Diehl:2005an} - this is the
amount of energy that must be given to a molecule to reach the fermionic scattering threshold. We stress the analogy
of eqs. (\ref{2bdConstraint}) to eqs. (\ref{CharPhases}). While the conditions (\ref{CharPhases}) describe a finite
temperature or classical phase transition, eqs.
(\ref{2bdConstraint}) imply a quantum phase transition from a fermionic vacuum to a molecular ground state
\cite{Diehl:2005an,Sachdev06}. Interestingly, the smooth crossover found at finite density terminates in a sharp phase
transition in the limit $k_F\to 0$.

Evaluating the bosonic mass term $\bar m_\phi^2$ in vacuum on the BEC side, one finds the well-known universal relation
between binding energy and scattering length in vacuum, $\epm = - 1/(Ma^2)$ in the broad resonance limit
$\hpb\to \infty$. This establishes the second order nature of the vacuum phase transition. For finite $\hpb$, scaling
violations $\mathcal O (\epm/\hpb^2)$ emerge \footnote{The situation is further complicated in the presence of an
additional scale set by a finite background scattering length \cite{Diehl:2006}.}. This gives a first glance at the
status of universality related to the value of $\hpb$.

\section{Universality}
\label{sec:Univ}

\begin{figure}[t!]
\begin{minipage}{\linewidth}
\begin{center}
\setlength{\unitlength}{1mm}
\begin{picture}(145,47)
      \put (25,0){
         \makebox(70,45){
     \begin{picture}(70,45)
      \put(0,0){\epsfxsize70mm \epsffile{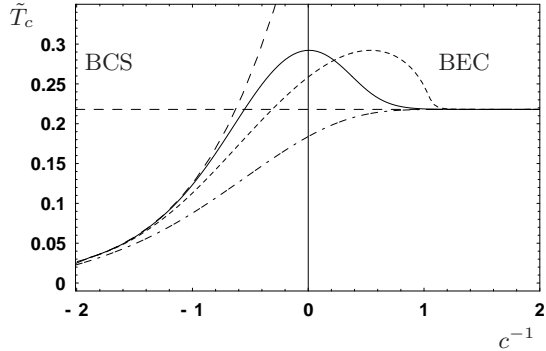}}
      \put(63,-3){$c^{-1}$}
      \put(-2,40){$\Tn_c$ }
      \put(8,34){BCS}
      \put(55,34){BEC}
      \end{picture}
      }}
\end{picture}
\end{center}
\vspace*{-1.25ex} \caption{Crossover phase diagram. The dependence of the critical temperature
$\tilde{T}_c = T_c/\epsilon_F$ on the inverse concentration $c^{-1}=(a k_F)^{-1}$ is shown for
the broad resonance limit $\hpn\to \infty$ (solid
and short-dashed line) and for the narrow resonance limit $\hpn \to 0$ (dashed-dotted line). We also indicate the standard
BEC (dashed horizontal line) and BCS (dashed rising line) limits which do not depend on the choice of the Yukawa coupling.
For the broad resonance limit we plot two different approximations as discussed in the text.}
\label{PhaseDiag}
\end{minipage}
\end{figure}

Universality refers to the ``loss of memory'' concerning details of the microscopic physics of a system.
A prominent example, which is often identified with the notion of universality, is the universal long-range behavior
of thermodynamic systems close to a classical or temperature-driven second order phase transition.
Here we use the idea of universality in a wider sense: Aspects of this phenomenon can be found in the thermodynamic
system when considering specific parameter regimes in the cube of scales fig. \ref{CubeOfScales}. We refer to them
as ``enhanced universality''. An example is provided by the BCS and BEC limits: here
the scale $|c^{-1}| \to \infty$ drops out. Physically, those regimes correspond to a loss of memory of field degrees of
freedom. This is particularly interesting in the BEC regime in the additional limit where there are no
classical bosonic degrees of freedom (broad resonance limit $\hpn\to \infty$, cf. below). Nevertheless, probed on
thermodynamic scales, the system precisely behaves as a gas of ``fundamental'' bosons described by a Bogoliubov-type theory
as stated in sect. \ref{FtConstr} and worked out in more detail in \cite{Diehl:2005an}. The reason is the formation of a two-body bound state as discussed in the last section,
which extends over much smaller distances $\sim a$ than the typical thermodynamic scales $k_F^{-1}$ or $\lambda_{dB}=
\sqrt{2\pi/ M T}$ in the BEC regime determined by $c^{-1} = (a k_F)^{-1} \gg 1$. The approach of the BCS and BEC regimes
can be seen in our result for the phase diagram fig. \ref{PhaseDiag}. In the following, we focus on an additional
aspect of universality related to the size of the Feshbach coupling $\hpn$ \cite{Diehl:2005an}.

\subsection{Exact Narrow Resonance Limit}
\label{sec:decoupling}

\begin{figure}[t!]
\begin{minipage}{\linewidth}
\begin{center}
\setlength{\unitlength}{1mm}
\begin{picture}(145,47)
      \put (25,0){
       \makebox(70,45){
    \begin{picture}(70,45)
      \put(0,0){\epsfxsize70mm \epsffile{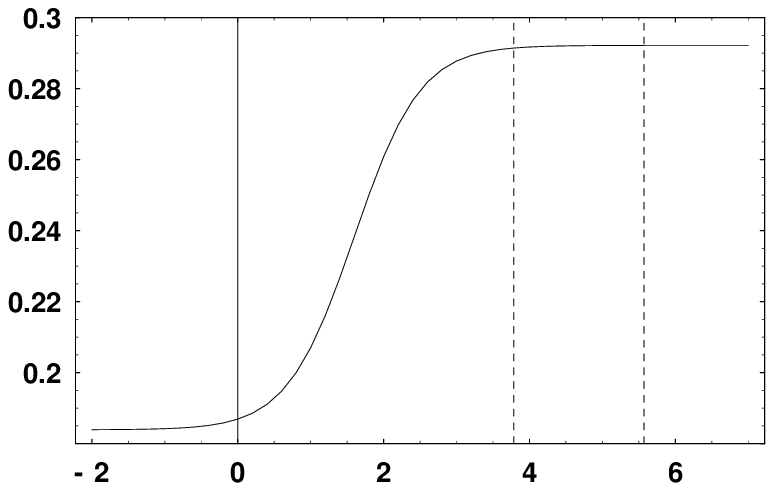}}
      \put(61,-3){$\log_{10}\hpn^2$}
      \put(40,25){$\kal$ } \put(59,25){$\lit$ }
      \put(-2,40){$\Tn_c$ }
      \end{picture}
           }}
\end{picture}
\end{center}
\vspace*{-1.25ex} \caption{ Enhanced universality for large and small $\hpn$. For $c^{-1} =0$ we plot the dependence on
$\hpn$ of the dimensionless critical temperature $\Tn_c$.
For small $\hpn < 1$, a stable universal narrow resonance limit is approached. For large $\hpn$ we find a very pronounced
insensitivity of $\Tn_c$ to the precise value of $\hpn$ -- note that we plot on a logarithmic scale. The
``crossover'' regime interpolates smoothly between the universal limits. The dashed lines correspond to the actual
value of $\hpn^2$ for $\lit$ and $\kal$. Indeed they belong to the class of broad
resonances.}
\label{hpUniversality}
\end{minipage}
\end{figure}

A nontrivial exact limit exists for which $\tilde{h}_\phi\to 0$ while $c$ and $\Tn$ are kept fixed. It applies to the
symmetric phase including the location of the critical line. This exact limit remains valid for arbitrary coupling $c$,
even if the scattering length $a$ is arbitrarily large. The existence of this limit guarantees that an appropriate
mean field theory remains valid as long as $\hpn$ remains small, say $\hpn <1$. This is confirmed in fig.
\ref{hpUniversality}. Nevertheless, our limit can describe the full BCS-BEC crossover as visible from the phase
diagram in fig. \ref{PhaseDiag} (dashed lower line).

We can project on the narrow resonance situation by considering the constrained limit $\hpn \to 0, c^{-1} =
-8\pi(\tilde{\nu} - 2\sigex)/\hpn^2 = \mathrm{const.}$ \footnote{We have set $\lambda_\psi = 0$.}. We may expand the
functional determinant in powers of the fluctuation $\delta\phi, \delta\phi^*$, which is now controlled by the smallness
of $\hpn$. The zero order contribution yields the fermionic mean field potential \footnote{$c^{-1} =\mathrm{const.}$
implies a fixed squared gap parameter $\rex = \hpn^2\tilde\phi^*\tilde\phi$ \cite{Diehl:2005an}.}. The remaining functional integral
is Gaussian and can be solved in one step, since higher powers in $\delta \phi$ also involve higher powers in $\hpn$.
The molecules and fermionic atoms decouple such that the Gaussian (one-loop) approximation becomes exact. The narrow
resonance limit is sensitive to further details of the microscopic physics, such as the classical gradient coefficient
$\bar A_{\phi,0}$. For fig. \ref{hpUniversality}, we have used $\bar A_{\phi,0}= 1/4M$ as obtained from a simple
symmetry consideration. It is universal in view of the insensitivity to the precise value of $\hpn$ as visible from fig.
\ref{hpUniversality}.

The narrow resonance limit describes a situation with large effective range $r_s =2 \bar A_{\phi,0}/\hpb^2$.
To see this, we rescale the field $\phi \to \varphi = \hpn \phi$ and perform the limit $\hpn \to 0$ at fixed
$\varphi$. This leads to a large weight in the exponential
\begin{eqnarray}
\sim \exp - \int \frac{r_s}{2} \,\,\hat\varphi^* \vec q\,^2 \hat \varphi, \qquad
r_s  \to \infty,
\end{eqnarray}
making the remaining functional integral Gaussian. The large weight controls the approximation for arbitrary $c^{-1}$
and in particular remains valid close to the resonance. The functional integral formulation hence shows the mechanism
controlling the approximation in the narrow resonance limit in a very clear way. The ordering principle emerging here
has effectively been used in \cite{Schwenk05}.

\subsection{Broad Resonance Limit}
\label{sec:broadres}

The broad resonance limit obtains for $\hpn \to \infty$ while again keeping $c$ fixed. It corresponds to a model
for fermionic atoms with local interaction and without explicit molecule degrees of freedom. For $\hpn\to \infty$ all
quantities depend only on $c$ and $\Tn$. The broad resonance limit therefore shows a particularly high degree of
universality.

We can qualitatively understand the mechanism for the broad resonance limit by again considering a fixed rescaled field
$\varphi = \hpn \phi$, now performing the limit $\hpn\to \infty$. After UV renormalization, this leads to a classical
mass term for the bosons $\sim c^{-1} = -8\pi(\tilde{\nu} - 2\sigex)/\hpn^2$. This ratio is kept fixed by requiring
$\tilde\nu \sim
\hpn^2$. At the same time, the kinetic coefficients in the propagator for the rescaled field vanish $\sim \hpn^{-2}$ -
the interaction becomes pointlike and the closed channel molecules reduce to purely auxiliary fields with no direct
physical meaning. Similarly, possible microscopic self-interactions of the closed channel molecules or higher order
interactions with the fermions are suppressed by powers of $\hpn^{-2}$. This demonstrates the
particularly high degree of universality for broad resonances -- it is extremely insensitive to microphysical details
concerning the closed channel molecules and uniquely determined by $c^{-1}$ and $\Tn$. Broad resonance universality
is stronger than the universality found for narrow resonances, which only referred to the insensitivity w.r.t.
the Feshbach coupling itself. Furthermore, the above consideration establishes the equivalence of
a purely fermionic setup as discussed by Strinati \emph{et al.} \cite{AAAAStrinati,CCStrinati} (single channel model)
and the two-channel model in the limit $\hpn\to \infty$. A complementary justification from a renormalization group
perspective, anticipated in \cite{Diehl:2005an}, has been given more recently in \cite{Sachdev06}.

The results for the broad resonance limit presented here still involve quantitative uncertainties. The shortcomings are
most severe in the crossover regime due to the absence of an obvious ordering principle. This is reflected in
the phase diagram fig. \ref{PhaseDiag}. The solid line results from an omission of boson
fluctuations in the Schwinger-Dyson equation from fig. \ref{SDEGraphic}, while they are included for the dashed line. Boson
fluctuations beyond Mean Field are subleading in the BCS and BEC regimes, where an ordering principle is provided by
large size of the classical bosonic mass term $\sim c^{-1}$. Substantial quantitative improvements can be obtained in
the frame of Functional Renormalization Group equations \cite{SDHGJPCW}.

In conclusion, we point out that a further ``crossover problem'', i.e. the crossover from small to large
$\hpn$ or from narrow to broad resonances, emerges from the above discussion. Physically, it describes the crossover from
nonlocal to pointlike interactions. Our interpretation of the transition from narrow to broad resonances in terms of a
decreasing effective range is confirmed in a numerical study \cite{Jensen04} comparing different types of interaction
potentials.

\subsection{Deviations from Universality}
\label{sec:DeviatUniv}

In the last section we focused on the universal aspects associated to the strict limits $\hpn\to 0, \hpn\to \infty$.
Relaxing the last condition to a finite but large $\hpn$, one expects small deviations from complete
universality, as anticipated in sect. \ref{sec:VacLim} for the vacuum. Here we will see that such scaling
violations can actually be probed experimentally. The systems currently investigated are
$\lit$ and $\kal$. Both range in the broad resonance regime, as can be seen from fig. \ref{hpUniversality}.

\begin{figure}[t]
\begin{minipage}{\linewidth}
\begin{center}
\setlength{\unitlength}{1mm}
\begin{picture}(105,53)
      \put (15,0){
     \makebox(100,50){
     \begin{picture}(100,50)
      \put(0,0){\epsfxsize80mm \epsffile{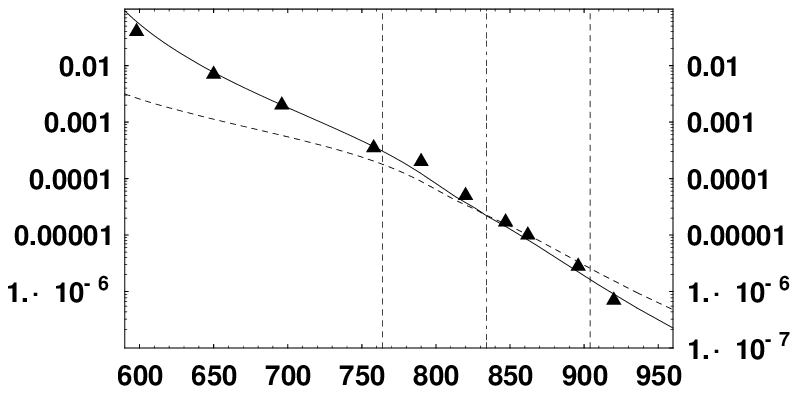}}
      \put(65,2){$B [\mathrm{G}]$}
      \put(-2,41){$\OmB$ }
       \put(0,20){{\large \textcolor{white}{$\blacksquare$} }}
       \put(0,17){{\large \textcolor{white}{$\blacksquare$} }}
       \put(3,17){{\large \textcolor{white}{$\blacksquare$} }}
        \put(87,17.5){{\large \textcolor{white}{$\blacksquare$} }}
       \put(87,12){{\large \textcolor{white}{$\blacksquare$} }}
        \put(87,9){{\large \textcolor{white}{$\blacksquare$} }}
         \put(85,17.5){{\large \textcolor{white}{$\blacksquare$} }}
       \put(85,12){{\large \textcolor{white}{$\blacksquare$} }}
        \put(85,9){{\large \textcolor{white}{$\blacksquare$} }}
      \end{picture}
      }}
\end{picture}
\end{center}
\vspace*{-3.25ex}
\caption{Fraction of closed channel
molecules $\OmB = \OmC + \OmM$, compared to experimental values \cite{Partridge05}, for $T=0$ and $k_F= 0.493
\mathrm{eV}\hat{=}250\mathrm{nK}$. The strongly interacting region $c^{-1}<1$ is indicated by vertical lines, where
the center line denotes the position of the resonance. The dashed line omits renormalization effects associated to the
background scattering $a_{bg}$.}
\label{BareExpPart}
\end{minipage}
\end{figure}

In a recent experiment Partridge \emph{et al.} \cite{Partridge05} assess the fraction of closed channel or bare
molecules $\OmB = \bar n_B/n$. The laser probe couples directly to their total number, i.e.
$\OmB = \OmM + \OmC$ with $\OmM$ and $\OmC$ the connected and condensate parts of the bare boson number density. Our
Yukawa-type formalism yields the following explicit formula in terms of dressed or renormalized quantities,
\begin{eqnarray}
\OmB = \OmM + \OmC = Z_\phi^{-1} (\Omega_M + \Omega_C) .
\end{eqnarray}
The fraction of bare molecules is $\mathcal O (\tilde h_\phi^{-2})$ -- the overall effect is small as can be seen from
fig. \ref{BareExpPart}. It might be viewed as a scaling violation to the fraction of dressed molecules $\Omega_M +
\Omega_C$. The dressed quantities $\Omega_M, \Omega_C$ are very insensitive to the precise value of the
microscopic Feshbach coupling $\tilde h_\phi$.  Scaling violations are $\mathcal{O}(\tilde h_\phi^{-2})$ or less and can
consequently be neglected in a leading order treatment in $\tilde h_\phi^{-2}$. In other words, having solved
the universal broad resonance limit, a systematic expansion in $\tilde h_\phi^{-2}$ is feasible.

We note that the non-universal domain probed here is in turn quite sensitive to further additional microphysical
information. In order to match the experimental data, one has to carefully include renormalization effects induced by
the background scattering $a_{bg}$. Furthermore, the nonlinear relation between
the magnetic field (or $\tilde\nu$) and the inverse scattering length in the presence of $a_{bg}$ (cf. eq.
(\ref{InMediumScattLength})) has to be taken into account. This is implemented by suitable Schwinger-Dyson equations.

Besides representing an interesting probe for the limitations of universality, the experiment of Partridge \emph{et al.}
resolves a true many-body effect. Recalling that the physics of the vacuum only supports a two-body bound state on the BEC
side of the resonance (below the center line at $B= 834 \mathrm G$), the presence of a nonvanishing though small fraction
of closed channel molecules can only be attributed to many-body induced bound state formation. This aspect is also
stressed in \cite{IIChen05,Stoof05}.

\section{Conclusion}
\label{sec:Conclusion}

The functional integral approach to the crossover problem in ultracold fermionic gases allows to fully exploit the presence
of the global $U(1)$ symmetry. The equation of state for the particle number is constructed as the conserved Noether
charge of the nonrelativistic quantum field theory. The concept of dressed bosonic fields emerges naturally from this derivation.
The thermodynamic phases can be classified by a symmetry consideration as well. While the thermodynamic equilibrium state
exhibits the full symmetry of the effective action in the normal gas phase above $T_c$, it has a spontaneously broken
$U(1)$ symmetry for $T< T_c$, irrespective to the value of the scattering length or further microphysical information.
The associated massless bosonic mode gives rise to superfluidity. This clearly reveals the universality of the
condensation phenomenon. Contact to microphysical observables is made through an appropriate vacuum limit of the
effective action. In this limit the smooth crossover terminates in a sharp quantum phase transition, which finds
an interpretation as the formation of a molecular two-body bound state.

Here we use this framework to work out the universal aspects in the phase diagram for ultracold fermionic atoms.
Besides the crossover from BCS to BEC, for which we present the finite temperature phase diagram, we find an additional
crossover as a function of the dimensionless
Feshbach coupling $\tilde h_\phi = 2M k_F^{-1/2} \bar h_\phi$ from a narrow resonance regime $\hpn\to 0$
or a situation with large effective range to a broad resonance limit $\hpn\to \infty$ with pointlike interactions. Narrow
resonances are universal in the sense of a pronounced insensitivity to the precise value of the microscopic Feshbach
coupling, while further microphysical scales are visible in the thermodynamics of the system. In contrast,
universality is greatly enhanced in the broad resonance regime. Similar to the narrow resonance situation,
the scale set by the microscopic Feshbach coupling drops out for $\hpn\to \infty$ and the physics becomes effectively
independent of $\hpn$. However, further microscopic
information is now suppressed with $\tilde h_\phi^{-2}$ except for the scattering length, which becomes the
unique parameter characterizing the interaction. We quantitatively assess deviations from the broad resonance
universality by computing the closed channel molecule fraction throughout the crossover for $\lit$, which has been
measured experimentally.

\textbf{Acknowledgement} -- The author would like to thank C. Wetterich for his collaboration on the work presented here,
and to J. Pawlowski and P. Strack for useful comments on the manuscript. Furthermore support from a
fellowship provided by the organizers of the ECT* school is acknowledged.

%


\bibliographystyle{hunsrt}
\bibliography{Citations}


\printindex
\end{document}